\documentclass[11pt,twoside]{article}

\usepackage{asp2006}
\usepackage{epsf}

\begin{document}
\title{Bars in field and cluster galaxies at intermediate redshifts}  
\author{Fabio D. Barazza\altaffilmark{1},  Pascale Jablonka\altaffilmark{1,2},
      and the EDisCS collaboration}   
\altaffiltext{1}{Laboratoire d'Astrophysique, EPFL, Observatoire de
  Sauverny, CH-1290 Versoix, Switzerland}
\altaffiltext{2}{Universit\'e de Gen\`eve, Observatoire de Sauverny,
  CH-1290 Versoix, Switzerland}

\begin{abstract} 
  We present the first study of large-scale bars in clusters at
  intermediate redshifts ($z=0.4-0.8$). We compare the properties of
  the bars and their host galaxies in the clusters with those of a
  field sample in the same redshift range. We use a sample of 945
  moderately inclined disk galaxies drawn from the EDisCS project. The
  morphological classification of the galaxies and the detection of
  bars are based on deep $HST/ACS$ $F814W$ images. The total optical
  bar fraction in the redshift range $z=0.4-0.8$, averaged over the
  entire sample, is $25\%$. This is lower than found locally, but in
  good agreement with studies of bars in field environments at
  intermediate redshifts. For the cluster and field subsamples, we
  measure bar fractions of $24\%$ and $29\%$, respectively. In
  agreement with local studies, we find that disk-dominated galaxies
  have a higher bar fraction than bulge-dominated galaxies. We also
  find, based on a small subsample, that bars in clusters are on
  average longer than in the field and preferentially found close to
  the cluster center, where the bar fraction is somewhat higher than
  at larger distances. 
\end{abstract}


\section{Introduction}   
Bars are believed to be very important with regard to the dynamical
and secular evolution of disk galaxies, particularly in redistributing
the angular momentum of the baryonic and dark matter components of
disk galaxies. Theory and $n$-body simulations predict that this
redistributions is characterized by the transfer of angular momentum to
the outer disk. As a result, gas is driven inside the corotation
radius toward the center of the disk, which can trigger starbursts
\citep{sak99,bou02,jog05} and contribute to the formation of disky
bulges \citep{kor04}.

While it is still unknown why a specific disk galaxy hosts a bar and
an apparently similar galaxy is unbarred, it is clear that a
significant fraction of bright disk galaxies appears barred in optical
observations \citep{esk00,ree07,mar07,bar08}. This result is mainly
based on samples of disk galaxies in field environments, whereas
studies of bars in cluster galaxies are rather rare. In general it is
found that the fraction of barred disks in clusters or groups is
roughly the same as in the field \citep{van07,mar09}, suggesting that
the denser environment does not significantly affect bar formation. On
the other hand, \cite{tho81} and \cite{and96} present evidence that
barred galaxies in the Coma and Virgo clusters are more concentrated
toward the cluster centers than unbarred disks.

We present results of the first study of bars in clusters at
intermediate redshifts and investigate the impact of the cluster
environment on bar formation and evolution. We use a sample of disk
galaxies from the ESO distant cluster survey
\citep[EDisCS,][]{whi05}. Using the available $I$-band $HST/ACS$
images we identify and characterize bars, based on quantitative
criteria. We look for relations between barred and unbarred galaxies
and their environment for a subsample, for which spectroscopic
redshifts and reliable cluster membership determinations are available.

\section{Sample and method}   
The EDisCS project has assembled three-band optical VLT deep
photometry, deep NTT/SOFI near-infrared imaging, and optical VLT/FORS2
spectroscopy for 26 optically selected and spectroscopically confirmed
galaxy structures between redshifts 0.39 and 0.96
\citep{hal04,mil08}. Additional HST/ACS images in the $F814W$ filter were
acquired for 10 fields containing the most distant clusters. Galaxies
in these 10 fields, regardless of whether they are cluster members or
group/field galaxies, and with $I<23$ mag constitute our basic sample.
From this sample we select all galaxies with Hubble types S0--Sm/Im based on
visual classification \citep{des07} and in the redshift range
$z=0.4-0.8$, which ensures to remain in the rest-frame optical (1906
objects). Results based on the separation between cluster and field
galaxies are based on a subsample, for which spectroscopic redshifts
and therefore a reliable cluster or field allocation is available (459
objects).

Our method to find bars relies on the fact that the isophotes of bars
in moderately inclined disk galaxies (i.e. with disk inclination
$i<60^{\circ}$) have much higher ellipticities than the isophotes of
the underlying disk. The ellipticities of the isophotes are derived
by fitting ellipses to the surface brightness distribution of the
disks using the IRAF task 'ellipse'. The corresponding profiles of
ellipticity ($\epsilon$) and position angle (P.A.) are investigated
based on two quantitative criteria: (1) $\epsilon$ increases steadily
to a global maximum higher than 0.25, while the P.A. value remains
constant (within $10^{\circ}$), and (2) $\epsilon$ then drops by at
least 0.1 and the P.A. changes at the transition between the bar to
the disk region. Galaxies meeting these two criteria have been
classified as barred.

\section{Results}
The optical bar fraction of the entire sample (including field and
cluster galaxies) is $\sim25\%$. This is significantly lower than is
typically found in optical studies of $local$ galaxies
\citep{esk00,ree07,mar07,bar08}, but in good agreement with studies of
galaxies in field environments at intermediate redshifts
\citep{jog04,elm04,sht08}. For the spectroscopically confirmed cluster
sample, we obtain $\sim24\%$, and for the corresponding field sample,
we derive $\sim29\%$. These values agree within the uncertainties with
the result for the complete sample and indicate that the frequency of
bars in clusters is almost identical to that in the field.
\begin{figure}[!t]
\plotone{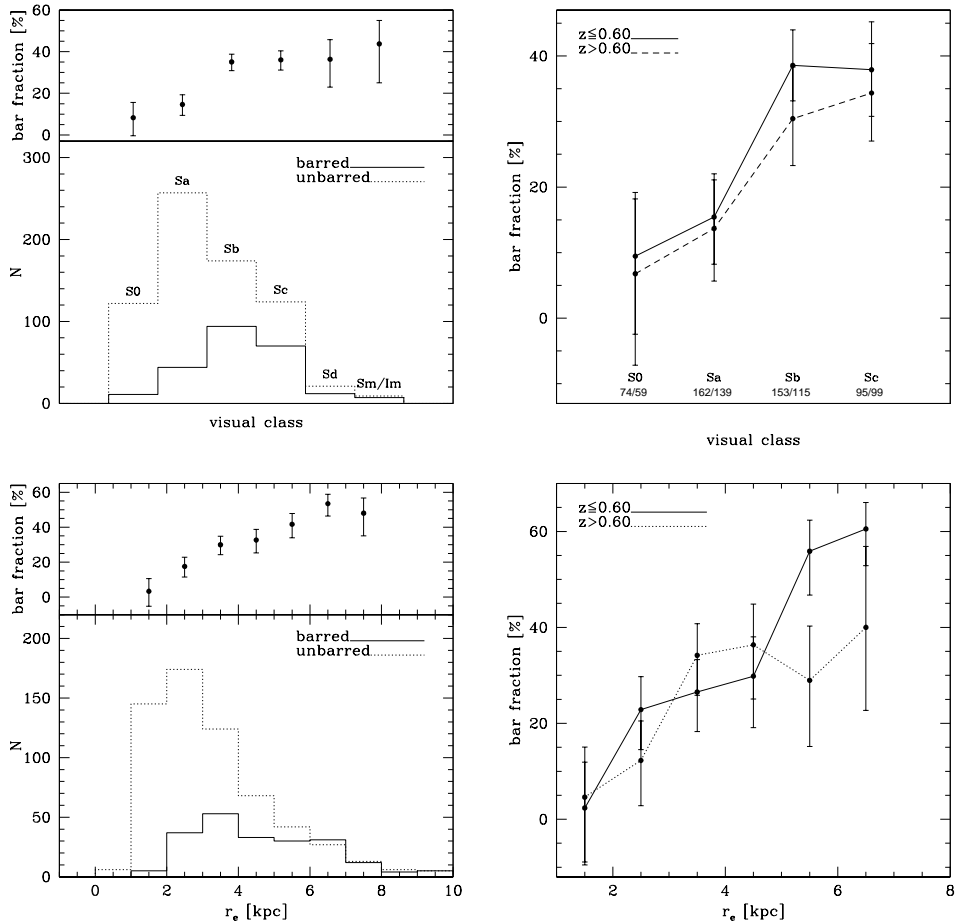}
\caption{$top$ $left:$ The number of barred and unbarred galaxies in the
  entire sample as a function of Hubble type and the corresponding bar
fraction. $top$ $right:$ The bar fraction as a function of Hubble type
for two redshift bins. $bottom$ $left:$ The number of barred and
unbarred galaxies in the entire sample as a function of effective
radius ($r_e$). $bottom$ $right:$ The bar fraction as a function of
$r_e$ for two redshift bins.}
\end{figure}
Figure 1 shows the optical bar fraction of the entire sample as a
function of Hubble type (top left) and effective radius ($r_e$,
bottom left). The effective radius defines the area, which contains
half of the total galaxy light. The bar fraction increases towards
later Hubble types and disks with larger effective radii. This
indicates that disk-dominated galaxies are more likely to be barred
than bulge-dominated galaxies. This results is in good agreement with
two recent SDSS studies also based on Hubble types and effective radius
\citep{bar08,agu09}. The right panels of Figure 1 show the same as the
left panels, but separated into a low and high redshift bin. The
relations remain roughly the same showing that the relative bar
fractions of the different Hubble types do not change significantly
with look-back time.
\begin{figure}
\plotone{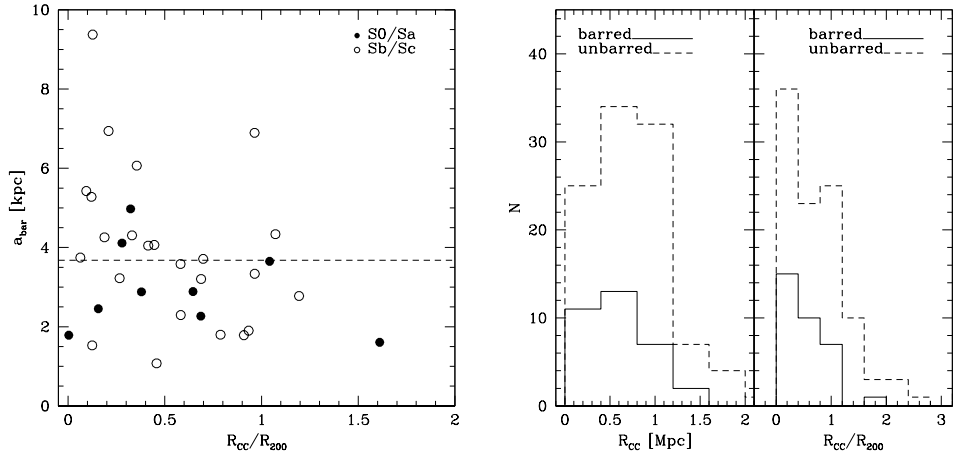}
\caption{$left:$ The bar size as a function of normalized
  clustercentric distance for the spectroscopic cluster
  subsample ($R_{200}$ is a measure of the virial radius). The dashed
  line indicates the mean bar size for this sample of 3.68
  kpc. $right:$ The distribution of the disk galaxies with respect to
  the clustercentric distance (left) and the normalized clustercentric
  distance (right) for the spectroscopic subsample.}
\end{figure}
The distribution of the barred galaxies within the clusters shows two
interesting features represented in Figure 2. The left panel shows
that the largest bars are preferentially found close to the cluster
centers and the right panel indicates that the bar fraction is
somewhat larger near the cluster centers than at larger radii. For the
$R_{CC}$ distribution, the bar fraction declines from {\bf $31\%$} in
the central bin to $18\%$ at $\sim1$ Mpc. For the $R_{CC}/R_{200}$
distribution, the corresponding values are {\bf $29\%$} in the central
bin and $22\%$ at $R_{200}$. We have to stress though that the sample
used is rather small, but we can safely say that barred galaxies do
not avoid the cluster center.

The question whether internal or external factors are more important
for bar formation and evolution cannot be answered definitely. On the
one hand, the bar fraction and properties of cluster and field samples
of disk galaxies are quite similar, indicating that internal processes
are crucial for bar formation. On the other hand, we find evidence
that cluster centers are favorable locations for bars, which suggests
that the internal processes responsible for bar growth are supported
by the typical interactions taking place in such environments.





\begin{thebibliography}{}
\bibitem[Andersen(1996)]{and96} Andersen, V.\ 1996, \aj, 111, 1805
\bibitem[Aguerri et al.(2008)]{agu09} Aguerri, J.~A.~L., Mendez-Abreu, J.,
Corsini, E.~M.,  2009, \aap, in press
\bibitem[Barazza et al.(2008)]{bar08} Barazza, F.~D., Jogee, S., \&
  Marinova, I.\ 2008, \apj, 675, 1194 
\bibitem[Bournaud \& Combes(2002)]{bou02} Bournaud, F., \& Combes, F.\
  2002, \aap, 392, 83 
\bibitem[Desai et al.(2007)]{des07} Desai, V., et al.\ 2007, \apj, 660, 1151
\bibitem[Elmegreen et al.(2004)]{elm04} Elmegreen, B.~G., Elmegreen, D.~M.,
\& Hirst, A.~C.\ 2004, \apj, 612, 191
\bibitem[Eskridge et al.(2000)]{esk00} Eskridge, P.~B., et al.\ 2000, \aj, 119,
536
\bibitem[Halliday et al.(2004)]{hal04} Halliday, C., et al.\ 2004,
  \aap, 427, 397
\bibitem[Jogee et al.(2005)]{jog05} Jogee, S., Scoville, N., \& Kenney,
J.~D.~P.\ 2005, \apj, 630, 837
\bibitem[Jogee et al.(2004)]{jog04} Jogee, S., et al.\ 2004, \apjl, 615, L105
\bibitem[Kormendy \& Kennicutt(2004)]{kor04} Kormendy, J., \& Kennicutt, R.~C.,
Jr.\ 2004, \araa, 42, 603
\bibitem[Marinova \& Jogee(2007)]{mar07} Marinova, I., \& Jogee, S.\ 2007,
\apj, 659, 1176
\bibitem[Marinova et al.(2009)]{mar09} Marinova, I. et al.\ 2009,
  \apj, in preparation
\bibitem[Milvang-Jensen et al.(2008)]{mil08} Milvang-Jensen, B., et
  al.\ 2008, \aap, 482, 419
\bibitem[Reese et al.(2007)]{ree07} Reese, A.~S., Williams, T.~B., Sellwood,
J.~A., Barnes, E.~I., \& Powell, B.~A.\ 2007, \aj, 133, 2846 
\bibitem[Sakamoto et al.(1999)]{sak99} Sakamoto, K., Okumura, S.~K.,
  Ishizuki, S., \& Scoville, N.~Z.\ 1999, \apj, 525, 691
\bibitem[Sheth et al.(2008)]{sht08} Sheth, K., et al.\ 2008, \apj, 675, 1141
\bibitem[Thompson(1981)]{tho81} Thompson, L.~A.\ 1981, \apjl, 244, L43
\bibitem[van den Bergh(2007)]{van07} van den Bergh, S.\ 2007, \aj, 134, 1508
\bibitem[White et al.(2005)]{whi05} White, S.~D.~M., et al.\ 2005,
  \aap, 444, 365
\end{thebibliography}
\end{document}